\documentclass[a4paper,onecolumn,11pt]{quantumarticle}
\pdfoutput=1
\usepackage[utf8]{inputenc}
\usepackage[english]{babel}
\usepackage[T1]{fontenc}
\usepackage{hyperref}

\usepackage{graphicx}
\usepackage{amsmath, amssymb, bm, physics}
\DeclareMathOperator*{\argmax}{arg\,max}
\DeclareMathOperator*{\argmin}{arg\,min}

\usepackage{color}

\usepackage{tikz}
\usepackage{lipsum}
\usepackage{caption}
\usepackage{subcaption}
\usepackage[numbers,sort&compress]{natbib}
\usepackage[lined,boxruled]{algorithm2e}

\usepackage{array}
\newcolumntype{P}[1]{>{\raggedright\arraybackslash}p{#1}} 

\begin{document}

\title{Coreset Clustering on Small Quantum Computers}
\author{Teague Tomesh}
\email{ttomesh@princeton.edu}
\affiliation{Department of Computer Science, Princeton University, 35 Olden Street, Princeton, NJ 08540}
\author{Pranav Gokhale}
\affiliation{Department of Computer Science, University of Chicago, 5730 S. Ellis Avenue, Chicago, IL 60637}
\author{Eric R. Anschuetz}
\affiliation{MIT Center for Theoretical Physics, 77 Massachusetts Avenue, Cambridge, MA 02139}
\author{Frederic T. Chong}
\affiliation{Department of Computer Science, University of Chicago, 5730 S. Ellis Avenue, Chicago, IL 60637}
\date{30 Apr 2020}

\maketitle

\begin{abstract}
Many quantum algorithms for machine learning require access to classical data in superposition. However, for many natural data sets and algorithms, the overhead required to load the data set in superposition can erase any potential quantum speedup over classical algorithms. Recent work by Harrow introduces a new paradigm in hybrid quantum-classical computing to address this issue, relying on coresets to minimize the data loading overhead of quantum algorithms. We investigate using this paradigm to perform $k$-means clustering on near-term quantum computers, by casting it as a QAOA optimization instance over a small coreset. We compare the performance of this approach to classical $k$-means clustering both numerically and experimentally on IBM Q hardware. We are able to find data sets where coresets work well relative to random sampling and where QAOA could potentially outperform standard $k$-means on a coreset. However, finding data sets where \emph{both} coresets and QAOA work well---which is necessary for a quantum advantage over $k$-means on the entire data set---appears to be challenging.
\end{abstract}

\section{Introduction}
Machine learning algorithms for analyzing and manipulating large data sets have become an integral part of today's world. Much of the rapid progress made in this area over the past decade can be attributed to the increased availability of large data sets that machine learning algorithms can train on and advances in hardware which accelerate the training process, such as the GPU. The last decade has also witnessed the emergence of prototype quantum computers which have been implemented using various qubit technologies, including superconducting circuits, trapped ions, neutral atoms, and solid state devices~\cite{murali2019full, pino2020demonstration, watson2018programmable, arute2019quantum}. The intersection between the emergent machine learning and quantum computing fields has produced many new algorithms which promise further advances in data processing capabilities.

Quantum algorithms such as HHL for solving linear systems~\cite{harrow2009quantum} and Grover's algorithm for database search~\cite{grover1996fast} are known to achieve exponential and quadratic speedups over their classical counterparts, respectively. However, many quantum algorithms for machine learning, including HHL and Grover search, also assume the use of an input data model (e.g. quantum RAM~\cite{giovannetti2008quantum}) which allows them to easily load classical data onto the quantum processor. This model is currently unrealistic~\cite{arunachalam2015robustness}. Without access to a quantum RAM, which presents the state $\ket{\psi}$ on demand, we resort to using a quantum circuit generate the desired state $\ket{\psi}$. Unfortunately, as the size of classical data sets grows to millions or billions of data points, the time and space requirements necessary to load the data may erase any potential quantum speedup. 

Recent work by Harrow~\cite{harrow2020small} introduces a new paradigm of hybrid quantum-classical computing to address this issue. The main idea is to take a large classical data set $\bm{X}$ and use a classical computer, potentially aided by a small quantum processor, to construct a coreset: a smaller, weighted data set $(\bm{X}', w)$ which sufficiently summarizes the original data. If the coreset is small enough (but still a faithful representation of $\bm{X}$), we could hope to optimize under the coreset with a small quantum computer. Prior work has focused on finding coreset construction algorithms that allow machine learning models to train on the coreset while remaining competitive with models that are trained on the entire data set~\cite{bachem2017practical, huggins2016coresets, campbell2018bayesian}. In~\cite{harrow2020small}, Harrow proposes three new hybrid algorithms which cover a range of machine learning tasks including maximum a posteriori estimation, inference, and optimization.

In this paper we evaluate the first of these new algorithms and adapt it to noisy quantum computers. The general version of this algorithm takes a data set $\bm{X}$ and cost function $f$ as input, uses a classical computer to construct a coreset $(\bm{X}', w)$, and then uses a quantum optimization algorithm to perform maximum a posteriori estimation (\cite{harrow2020small}, Algorithm 1). A specific instance of the algorithm is also outlined which solves the $k$-means clustering problem (\cite{harrow2020small}, Algorithm 1.1). The specific case of $k$-means is the focus of this paper. At a high level, this algorithm solves $k$-means clustering on a data set $\bm{X}$ by first constructing a coreset $(\bm{X}', w)$, and then optimally clustering $(\bm{X}', w)$ with Grover search.

\begin{algorithm}[t]
\SetAlgoLined
\SetKwData{Left}{left}\SetKwData{This}{this}\SetKwData{Up}{up}
\SetKwFunction{Union}{Union}\SetKwFunction{FindCompress}{FindCompress}
\SetKwInOut{Input}{Input}\SetKwInOut{Output}{Output}\SetKwInOut{Algorithm}{Algorithm}
\Input{A data set $\bm{x}_1, ..., \bm{x}_n \in \mathbb{R}^d$}
\Output{Cluster centers $\bm{\mu}_{-1}, ..., \bm{\mu}_{+1}$ which approximately minimize
\begin{equation}
    \sum_{i \in [n]}\min_{j \in \{\bm{\mu}_{-1}, \bm{\mu}_{+1}\}}\left\lVert\bm{x}_i - \bm{\mu}_{j} \right\rVert^2
    \nonumber
\end{equation}}
\Algorithm{}
1. Construct a coreset $(\bm{X}', w)$\ of size $m$.\\
2. Construct a $j$th order $m$-qubit Hamiltonian for the coreset.\\
3. Use QAOA to variationally approximate an energy-maximizing eigenstate of the Hamiltonian.\\
4. Treat the 0/1 assignment of the eigenstate as the $k=2$ clustering.
 \caption{$2$-means clustering via coresets+QAOA}
 \label{alg:kmeans_qaoa}
\end{algorithm}


However, Grover search is unlikely to be tenable on noisy devices \cite{regev2008impossibility}. As proposed in \cite{harrow2020small}, we reformulate the coreset clustering problem as a Quantum Approximate Optimization Algorithm (QAOA)~\cite{farhi2014quantum} instance. QAOA is variationally optimized and is able to tolerate some noise when coupled with a robust classical optimizer. For simplicity, we study $2$-means clustering specifically. Algorithm~\ref{alg:kmeans_qaoa} summarizes our approach.

Our core contributions are as follows:
\begin{itemize}
    \item We implement algorithms for coresets and evaluate their performance on real data sets.
    \item We cast coreset clustering to a Hamiltonian optimization problem that can be solved with QAOA. We also demonstrate how to break past the assumption of equal cluster weights.
    \item We benchmark the performance of Algorithm~\ref{alg:kmeans_qaoa} across six different data sets including real and synthetic data. We compare the $2$-means clustering solutions found by coresets+QAOA with the solutions given by standard algorithms for solving $2$-means on both the full data sets and the coresets. We find that some data sets are better suited to coreset summarization than others which can play a large role in the overall performance of the algorithm.
\end{itemize}

In our evaluations, the size of the coresets constructed in step 1 of Algorithm~\ref{alg:kmeans_qaoa} are limited by the size of the quantum computer used in step 3. For some data sets, this restriction on coreset size negatively impacts the performance of clustering on the coresets when compared to $k$-means on the entire data set. Nonetheless, we are able to at least show cases where QAOA-based clustering on the coresets is competitive with the standard $2$-means algorithms on those coresets. This suggests that Algorithm~\ref{alg:kmeans_qaoa} will improve when quantum computers can support more qubits and more gates, thereby allowing them to utilize larger coresets. However, our evaluations also suggest that either high $m$ (and thereby many qubits) or a high order QAOA implementation (with many gates) will be needed for a possible quantum advantage on typical data sets.

The rest of the paper is organized as follows. In Section~\ref{sec:kmeans} we give an overview of the $k$-means clustering problem. Section~\ref{sec:coresets} discusses coresets for $k$-means. Section~\ref{sec:coreset_qaoa} describes the reduction from $k$-means to QAOA. We present and discuss our results in Sections~\ref{sec:results} and~\ref{sec:discussion}.

\section{$k$-means Clustering}\label{sec:kmeans}
The $k$-means clustering problem takes an input data set $\bm{x}_1, ..., \bm{x}_n \in \mathbb{R}^d$ and aims to identify cluster centers $\bm{\mu}_1, ..., \bm{\mu}_k$ that are near the input data. Typically, $k\ll n$; for simplicity we focus on $k=2$. Foreshadowing quantum notation, we will prefer to denote our cluster centers as $\bm{\mu}_{-1}$ and $\bm{\mu}_{+1}$. Then, the objective of this $2$-means problem is to find the partitioning of $[n]$ into two sets $S_{-1}$ and $S_{+1}$ that minimizes the squared-distances from the closest cluster centers:
\begin{equation} \sum_{i \in S_{-1}} \left\lVert\bm{x}_i - \bm{\mu}_{-1} \right\rVert^2 + \sum_{i \in S_{+1}} \left\lVert\bm{x}_i - \bm{\mu}_{+1} \right\rVert^2. \label{eq:2means} \end{equation}
While the cluster centers $\bm{\mu}_{-1}$ and $\bm{\mu}_{+1}$ appear to be free variables, it can be shown~\cite{bishop2006pattern} that they are uniquely determined by the $S_{-1}$ and $S_{+1}$ partitionings in order to minimize Eq.~\eqref{eq:2means}. In particular, these cluster centers are the centroids,
\begin{equation*}
    \bm{\mu}_{-1} = \frac{\sum_{i \in S_{-1}} \bm{x}_i}{\left\lvert S_{-1}\right\rvert} \quad \text{ and } \quad \bm{\mu}_{+1} = \frac{\sum_{i \in S_{+1}} \bm{x}_i}{\left\lvert S_{+1}\right\rvert} .
\end{equation*}

Thus, in principle the objective function in Eq.~\eqref{eq:2means} can be minimized by evaluating all $2^n$ possible partitionings of $[n]$ into $S_{-1}$ and $S_{+1}$. However, this brute force exponential scaling is impractical, even for modest $n$. Instead, $k$-means is typically approached with heuristics like Lloyd's Algorithm~\cite{lloyd1982least}, which does not guarantee optimal solutions in polynomial time, but performs well in practice. Moreover, relatively simple improvements to the initialization step in Lloyd's Algorithm lead to performance guarantees. Notably, the $k$-means++ initialization procedure guarantees $(8 \ln k + 2)$-competitive solutions in the worst case~\cite{arthur2006k}.

For many data sets, Lloyd's Algorithm augmented with $k$-means++ initialization rapidly converges to close-to-optimal (often optimal) solutions. However, in general, finding the optimal cluster centers is a computationally hard problem. Even for $k=2$, the problem is \textsc{NP}-hard~\cite{garey1982complexity, aloise2009np}.

\section{Coresets for $k$-means}\label{sec:coresets}

An $\epsilon$-coreset for $k$-means is a set of $m$ (typically $<< n$) weighted points such that the optimal $k$-means clustering on the coreset is within $(1+\epsilon)$ of the optimal clustering on the entire data set of $n$ points. A coreset data reduction is appealing because we would prefer to solve a problem over $m << n$ points. The size $m$ needed depends on the target error $\epsilon$, $k$, the data set dimension $d$, and the probability of success $\delta$. We implemented two coreset procedures. The first, Algorithm 2 of \cite{bachem2017practical}, gives a coreset size of $m = O(\frac{d k^3 \log{k} + k^2 \log{\frac{1}{\delta}}}{\epsilon^2})$. The second, Algorithm 2 of \cite{braverman2016new}, gives a coreset size of $m = O(\epsilon^{-2} k \log{k} \min(\frac{k}{\epsilon}, d))$.

One might hope to pick a target $\epsilon$ and then pick $m$ accordingly. However, the exact expressions---including constants---for the scaling of $m$ are not readily available. Regardless, our goal is simply to probe the limits of small current-generation quantum computers, which have at most a few dozen qubits. Therefore, we approach coreset construction in the reverse direction by first choosing $m$ and then evaluating the performance of the resulting coreset. As discussed in the next section, $m$ will equal the number of qubits we need. Therefore, we choose $m \in \{5, 10, 15, 20\}$ for our evaluations.

For implementations of the coreset algorithms in both, \cite{bachem2017practical} and \cite{braverman2016new}, an $(\alpha, \beta)$ bicriterion approximation is required. We use $D^2$ sampling, which is the initialization step for $k$-means++ \cite{arthur2006k}, as our bicriterion approximation. We chose $\beta = 2$, which corresponds to picking $\beta k = 4$ centroids in the bicriterion approximation. For each data set, we selected the best (lowest cost) approximation from 10 repeated trials, as is also done by Scikit-learn's default implementation of $k$-means.

Through our evaluations, we did not find significant differences between the performances of the \cite{bachem2017practical} and \cite{braverman2016new} coreset algorithms. In fact, we did not observe a significant improvement over random sampling either, except for a synthetic data set with a few rare and distant clusters.

\section{Coreset $k$-means via QAOA} \label{sec:coreset_qaoa}

\subsection{QAOA}

The Quantum Approximate Optimization Algorithm (QAOA)~\cite{farhi2014quantum} is a quantum variational algorithm inspired by the quantum adiabatic algorithm~\cite{farhi2000quantum}. The adiabatic theorem implies that, for large enough $T$, starting in the $\ket{+}^{\otimes m}$ state and performing time-evolution under the time dependent Hamiltonian:
\begin{equation*}
    H\left(t\right)=\left(1-\frac{t}{T}\right)\sum\limits_{i=1}^m X_i+\frac{t}{T} H_P
\end{equation*}
from $t=0$ to $t=T$ results in a state with high overlap with the $m$ qubit state $\ket{z_{\text{sol}}}$, where
\begin{equation*}
    \ket{z_{\text{sol}}}=\argmax_{\ket{z}}\bra{z}H_P\ket{z}.
\end{equation*}
For concreteness we assume that $H_P$ is diagonal such that $\ket{z_{\text{sol}}}$ is a computational basis state. One can approximate this adiabatic evolution with a finite Trotterized evolution
\begin{equation}
    \ket{z_{\text{sol}}}\approx \ket{\bm{\beta},\bm{\gamma}}\equiv\prod_{j=1}^p \mathrm{e}^{-\mathrm{i}\beta_j H_M}\mathrm{e}^{-\mathrm{i}\gamma_j H_P}\ket{+}^{\otimes m}
    \label{eq:qaoa_ansatz}
\end{equation}
for certain $\bm{\beta},\bm{\gamma}$, where $H_M=\sum\limits_{i=1}^m X_i$. In the limit $p\to\infty$, the Trotter decomposition and the adiabatic theorem imply that there exist $\bm{\beta}$ and $\bm{\gamma}$ such that this approximation is exact; \textit{a priori}, however, it is not obvious what one should choose for these parameters for finite $p$ to tighten the approximation in Eq.~\eqref{eq:qaoa_ansatz}. Therefore, QAOA combines the ansatz of Eq.~\eqref{eq:qaoa_ansatz} with a classical optimization loop, performing the maximization of the function
\begin{equation*}
    F\left(\ket{\bm{\beta},\bm{\gamma}}\right)=\bra{\bm{\beta},\bm{\gamma}}H_P\ket{\bm{\beta},\bm{\gamma}}.
\end{equation*}
By the variational principle, for large enough $p$ the $\argmax$ of this optimization will approximate $\ket{z_{\text{sol}}}$. In practice, a quantum computer evaluates $F\left(\ket{\bm{\beta},\bm{\gamma}}\right)$ (or e.g. gradients of $F\left(\ket{\bm{\beta},\bm{\gamma}}\right)$), whilst a classical computer uses the function evaluations to heuristically optimize the function. In the remainder of this Section we describe how one can interpret the solution of the $k$-means problem as the highest excited state of a diagonal Hamiltonian, which can be heuristically found using QAOA. We note that prior work in \cite{otterbach2017unsupervised} also proposed and experimentally demonstrated clustering via QAOA instance; our work and reduction is more specific to $k$-means clustering.

\subsection{Hamiltonian Interpretation of $k$-means: Equal Cluster Weights} \label{subsec:equal_weights}

Under the weighted vectors of a coreset of size $m$, the $2$-means objective function is similar to that of Eq.~\eqref{eq:2means}, but now each input vector $\bm{x}_i$ has an associated weight $w_i$. The modified objective function is then \begin{equation} \sum_{i \in S_{-1}} w_i \left\lVert\bm{x}_i - \bm{\mu}_{-1} \right\rVert^2 + \sum_{i \in S_{+1}} w_i \left\lVert\bm{x}_i - \bm{\mu}_{+1} \right\rVert^2, \label{eq:2means_weighted} \end{equation}
where the cluster centers are now also weighted such that:
\begin{equation*}
    \bm{\mu}_{-1} = \frac{\sum_{i \in S_{-1}} w_i \bm{x}_i}{W_{-1}} \quad \text{ and } \quad \bm{\mu}_{+1} = \frac{\sum_{i \in S_{+1}} w_i \bm{x}_i}{W_{+1}}.
\end{equation*}
Here, $W_{\pm 1} = \sum_{i \in S_{\pm 1}} w_i$. We also define $W \equiv W_{-1} + W_{+1}$.

As shown in Appendix~\ref{app:reduction}, minimizing Eq.~\eqref{eq:2means_weighted} is equivalent to maximizing the weighted intercluster distance
\begin{equation}
    W_{+1}W_{-1} \left\lVert \bm{\mu}_{+1} - \bm{\mu}_{-1} \right\rVert^2. \label{eq:weighted_cluster_distance}
\end{equation}
In this section, we consider the case where the optimal clusters have equal weights, $W_{+1} = W_{-1}$. Often, this is a good approximation, because $W_{+1}W_{-1}$ is maximized for $W_{+1} = W_{-1} = W/2$.

Under this constraint, Eq.~\eqref{eq:weighted_cluster_distance} reduces to
\begin{equation*}
\begin{split}
    &\left(\frac{W}{2} \right)^2 \left\lVert  \frac{\sum_{i \in S_{-1}} w_i \bm{x}_i}{W/2} - \frac{\sum_{i \in S_{+1}} w_i \bm{x}_i}{W/2} \right\rVert^2 
    = \left\lVert \sum_{i \in S_{-1}} w_i \bm{x}_i - \sum_{i \in S_{+1}} w_i \bm{x}_i \right\rVert^2 \\
    &= \sum_{i} w_i^2 \left\lVert\bm{x}_i\right\rVert^2 + 2 \sum_{i < j \in S_{-1}} w_i w_j \bm{x}_i \cdot \bm{x}_j + 2 \sum_{i < j \in S_{+1}} w_i w_j \bm{x}_i \cdot \bm{x}_j-2 \sum_{i \in S_{-1}, j \in S_{+1}} w_i w_j \bm{x}_i \cdot \bm{x}_j \\
    &= \sum_{i} w_i^2 \left\lVert\bm{x}_i\right\rVert^2 + 2 \sum_{i < j} w_i w_j \bm{x}_i \cdot \bm{x}_j - 4 \sum_{i \in S_{-1}, j \in S_{+1}} w_i w_j \bm{x}_i \cdot \bm{x}_j.
\end{split}
\end{equation*}

In this expression, only the third term is dependent on our $S_{-1}, S_{+1}$ partitioning. Therefore, the $2$-means objective for equal cluster weights is equivalent to maximizing the (re-scaled) third term:

\begin{equation}
\begin{split}
    \sum_{i \in S_{-1}, j \in S_{+1}} - w_i w_j \bm{x}_i \cdot \bm{x}_j. \label{eq:final_unweighted}
\end{split}
\end{equation}

\begin{figure}[h]
    \centering
    \includegraphics[width=0.65\textwidth]{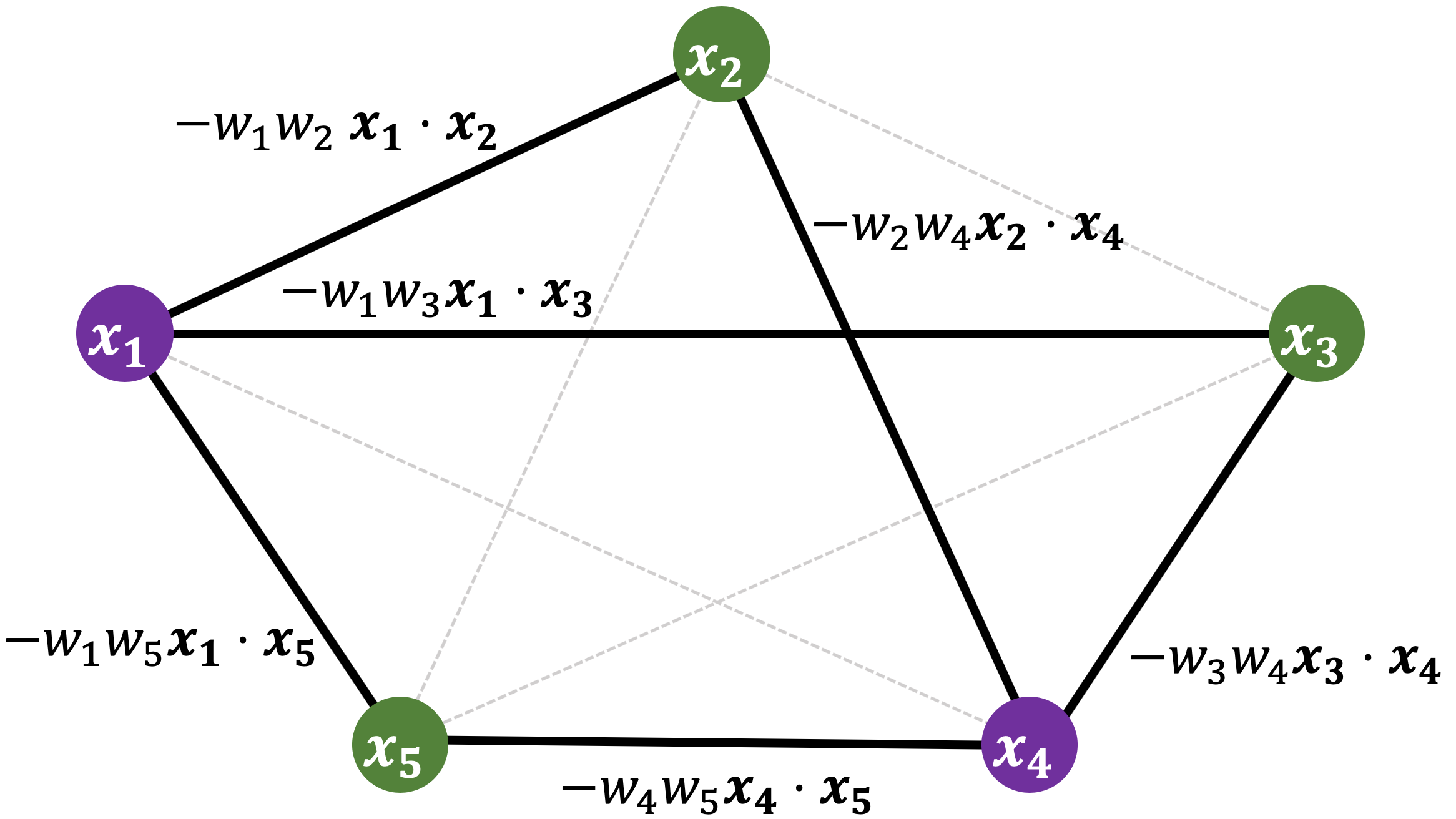}
    \caption{Weighted MAX-CUT for a coreset consisting of five points. Given an assignment of vertices to two colors, i.e. a cut, we are interested in the sum of $- w_i w_j \bm{x}_i \cdot \bm{x}_j$ on edges crossing the cut. By interpreting these terms as edge weights, we seek a weighted MAX-CUT.}
    \label{fig:weighted_maxcut}
\end{figure}

This expression can be interpreted as a weighted MAX-CUT problem on a complete graph. Each vertex in the graph represents a coreset point $\bm{x}_i$, and the weight of the $\left(i,j\right)$ edge in the graph is $-w_i w_j \bm{x}_i \cdot \bm{x}_j$. Our objective is to assign labels $\pm 1$ to each vertex in the graph such that the sum of edge weights over cut-crossing edges is maximized. Figure~\ref{fig:weighted_maxcut} depicts an example for a coreset containing five points. Although all 10 edges have weights, we only sum the weights of edges crossing the cut. For the particular coloring (partitioning) in Figure~\ref{fig:weighted_maxcut}, six of the edges cross the cut.

In order to maximize Eq.~\eqref{eq:final_unweighted} using QAOA, we must encode it as a Hamiltonian. For $Z_i, Z_j \in \{-1, +1\}$, note that $\frac{1 - Z_i Z_j}{2}$ is 1 for $Z_i \neq Z_j$ and 0 for $Z_i = Z_j$. Therefore, Eq.~\eqref{eq:final_unweighted} corresponds to the energy of the problem Hamiltonian:
\begin{equation*}
\tilde{H}_P = \frac{1}{2}\sum_{i < j} w_i w_j \bm{x}_i \cdot \bm{x}_j\left(Z_i Z_j - 1\right),
\end{equation*}
which is maximized at the same assignment of $\{Z_i\}$ as the offset-and-scaled Hamiltonian:
\begin{equation}
\begin{split}
H_P = \sum_{i < j} w_i w_j \bm{x}_i \cdot \bm{x}_j Z_i Z_j. \label{eq:problem_ham}
\end{split}
\end{equation}

\subsection{Hamiltonian Interpretation of $k$-means: Unequal Cluster Weights}

Now, we once again begin with Eq.~\eqref{eq:weighted_cluster_distance}:
\begin{equation*}
    W_{+1}W_{-1} \left\lVert\bm{\mu}_{+1} - \bm{\mu}_{-1}\right\rVert^2.
\end{equation*}

Unlike in Sec.~\ref{subsec:equal_weights}, however, we now longer assume that $W_{+1} = W_{-1} = W/2$. We can instead write:
\begin{equation}
\begin{split}
&W_{-1} W_{+1} \left\lVert\bm{\mu}_{-1} - \bm{\mu}_{+1}\right\rVert^2 = \left\lVert\frac{\sqrt{W_{-1} W_{+1}}}{W_{-1}} \sum_{i \in S_{-1}} w_i \bm{x}_i - \frac{\sqrt{W_{-1} W_{+1}}}{W_{+1}} \sum_{i \in S_{+1}} w_i \bm{x}_i\right\rVert^2 \\
&= \left\lVert\frac{\sqrt{W_{+1}}}{\sqrt{W_{-1}}} \sum_{i \in S_{-1}} w_i \bm{x}_i - \frac{\sqrt{W_{-1}}}{\sqrt{W_{+1}}} \sum_{i \in S_{+1}} w_i \bm{x}_i\right\rVert^2 = \sum_{i} \left( \frac{W_{+1}}{W_{-1}} \text{ if $i \in S_{-1}$, else } \frac{W_{-1}}{W_{+1}} \right) w_i^2 \left\lVert\bm{x}_i\right\rVert^2 \\
&+ 2 \sum_{i < j} \left( \frac{W_{+1}}{W_{-1}} \text{ if $i, j \in S_{-1}$,} \text{ } \frac{W_{-1}}{W_{+1}} \text{ if $i, j \in S_{+1}$, else} -1  \right) w_i w_j \bm{x}_i \cdot \bm{x}_j.
\end{split}
\label{eq:ratiod_eq}
\end{equation}

\begin{figure}[h]
    \centering
    \includegraphics[width=0.45\textwidth]{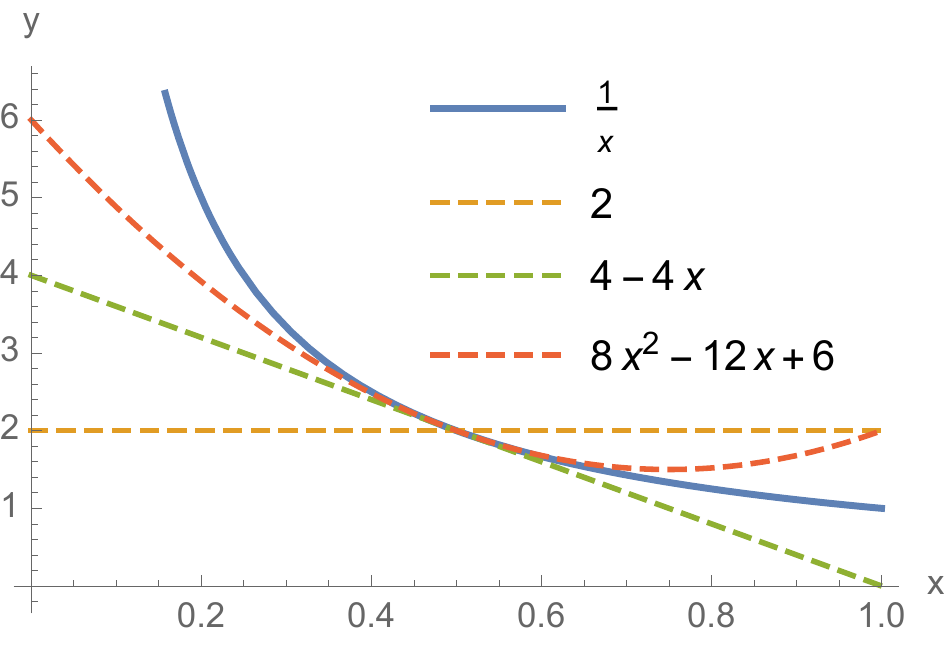}
    \caption{First three Taylor expansions of $y=\frac{1}{x}$ around $x=0.5$.}     \label{fig:taylor}
\end{figure}

We now examine the ratios
\begin{equation}
    \frac{W_{+1}}{W_{-1}} = \frac{1}{W_{-1}/W} - 1 \quad \text{and} \quad \frac{W_{-1}}{W_{+1}} = \frac{1}{W_{+1}/W} - 1.
    \label{eq:ratios}
\end{equation}
and we consider the Taylor expansion of this expression around $x\equiv W_{-1}/W = W_{+1}/W = 1/2$, i.e. around the equal cluster weight scenario of Sec.~\ref{subsec:equal_weights}. The motivation for using a Taylor expansion is that the resulting polynomial has operational significance as a Hamiltonian and can be written in terms of $Z_i$'s. $1/x$ and the three leading orders of its Taylor expansion are shown in Figure~\ref{fig:taylor}.

The zeroth order Taylor expansion, $1/x \approx 2$, corresponds to the case of equal cluster weights. Keeping terms through first order gives $4 - 4x$. Informally, this linear approximation is close to $1/x$ for $0.4 < x < 0.6$. Therefore, we could reasonably hope for the first order Taylor expansion to work well for slightly unbalanced optimal clusters, perhaps with 3:2 imbalances.

With this motivation, we expand Eq.~\eqref{eq:ratios} and find that:
\begin{equation*}
    \frac{W_{+1}}{W_{-1}} \approx 3 - \frac{4}{W} W_{-1} \quad \text{and}\quad \frac{W_{-1}}{W_{+1}} \approx 3 - \frac{4}{W} W_{+1}.
\end{equation*}

Performing a similar translation into Pauli $Z$ operators as in Sec.~\ref{subsec:equal_weights}, these terms can be expressed in Hamiltonian form and simplified to
\begin{equation*}
\frac{W_{+1}}{W_{-1}} \approx 1 + \frac{2}{W}\sum\limits_l w_l Z_l \quad \text{and} \quad \frac{W_{-1}}{W_{+1}} \approx 1 - \frac{2}{W} \sum\limits_l w_l Z_l.
\end{equation*}

Plugging these into Eq.~\eqref{eq:ratiod_eq} gives
\begin{equation*}
\begin{split}
&W_{-1} W_{+1} \left\lVert\bm{\mu}_{-1} - \bm{\mu}_{+1}\right\rVert^2 \\
& =\sum\limits_i \left( 1 + \frac{2}{W} \sum\limits_l w_l Z_l \text{ if $i \in S_{-1}$, else } 1 - \frac{2}{W} \sum\limits_l w_l Z_l \right) w_i^2 \left\lVert\bm{x}_i\right\rVert^2 \\
&+ 2 \sum_{i < j} \left( 1 + \frac{2}{W} \sum\limits_l w_l Z_l \text{ if $i, j \in S_{-1}$,} \text{ } 1 - \frac{2}{W} \sum\limits_l w_l Z_l \text{ if $i, j \in S_{+1}$, else} -1  \right) w_i w_j \bm{x}_i \cdot \bm{x}_j.
\end{split}
\end{equation*}

The indicator functions can also be rewritten as Pauli expressions, resulting in the final problem Hamiltonian:
\begin{equation}
\begin{split}
 \sum_{i} \left( 1 - \frac{2 Z_i}{W} \sum_l w_l Z_l \right) w_i^2 \left\lVert\bm{x}_i\right\rVert^2 + 2 \sum_{i < j} \left( Z_iZ_j - \frac{Z_i + Z_j}{W} \sum_l w_l Z_l \right) w_i w_j \bm{x}_i \cdot \bm{x}_j.
\label{eq:final_weighted_ham}
\end{split}
\end{equation}

Interestingly, the Eq.~\eqref{eq:final_weighted_ham} Hamiltonian only has quadratic terms, which is no more difficult to implement than the zeroth order case that assumes equal cluster weights. However, for higher order Taylor expansions, the degree of the Hamiltonian will increase; a second order Taylor expansion will have a cubic Hamiltonian, a third order Taylor expansion will have a quartic Hamiltonian, and so forth.

\section{Results}\label{sec:results}
\subsection{Data sets}
Table~\ref{tab:data sets} describes the six data sets we used in our evaluations. The Epilepsy, Pulsars, and Yeast data sets are part of the UCI Machine Learning Repository~\cite{Dua:2019}. For Common Objects in Context (COCO), the image pixels were preprocessed with the img2vec~\cite{safka2017img2vec} library. This library translates the pixels of each image into a 512-dimensional feature vector using a Resnet-18 model~\cite{he2016deep} pretrained on the ImageNet data set~\cite{deng2009imagenet}.

\begin{table}
\renewcommand{\arraystretch}{1.3} \centering
\begin{tabular}{P{0.12\textwidth}|P{0.78\textwidth}}
data set & Description \\ \hline
CIFAR-10 & 10k images (32x32 pixels) from CIFAR-10 data set~\cite{krizhevsky2009learning}. 1k images per category. \\
COCO & 5k images from Common Objects in Context validation data set~\cite{lin2014microsoft}. Images translated into feature vectors of dimension 512. \\
Epilepsy & Epileptic seizure recognition data set from~\cite{andrzejak2001indications}. 11.5k vectors of dimension 179. \\
Pulsars & Pulsar candidates from HTRU2 data set~\cite{lyon2016fifty}. 1.6k/17.9k of 9-dimensional feature vectors are pulsars. \\
Yeast & Localization sites of proteins~\cite{horton1996probabilistic}. 1.5k 8-dimensional vectors. \\
Synthetic & 40k 512-dimensional points drawn from 11 random Gaussian clusters. Ten clusters contribute 5 points each, last cluster has majority.
\end{tabular}
\caption{Data sets evaluated.}
\label{tab:data sets}
\end{table}

\subsection{Evaluation Methodology}

We evaluated $2$-means on each of these data sets using Scikit-learn \cite{pedregosa2011scikit}. We used Scikit-learn's default implementation of $k$-means, which initializes clusters with best-of-10 $k$-means++ \cite{arthur2006k} and terminates either after 300 iterations or upon stabilizing within $10^{-4}$ relative tolerance. This default implementation is an aspirational, though realistic, target against which to compare QAOA. The cost we report is the ``sum of squared distances to nearest cluster'' objective function in Eq.~\eqref{eq:2means_weighted}, also referred to as inertia~\cite{ralambondrainy1995conceptual}.

On each data set, we computed $m$= 5, 10, 20, and 40 uniformly random samples, as well as $m$-coresets using Algorithm 2 of~\cite{braverman2016new}. Then, we ran the 2-means clustering implementation on this coreset, and evaluated the cost of the output cluster centers, evaluated against the entire data set. For each data set, we ran this process 10 times. On five of the six data sets, we report best-of-10 results, since in practice one would indeed choose the best result. For the Synthetic data set, we report average, best, and worst costs, to emphasize that the coreset algorithm is consistently better than random sampling.

In addition to these classical results, we took the best $m=5$ and $m=10$ coresets and constructed Hamiltonians for them, as described in Sec.~\ref{sec:coreset_qaoa}. For $m=10$, we constructed Hamiltonians with zeroth order, first order, second order, and infinite order Taylor expansions. For $m=5$, we only constructed the zeroth order Hamiltonian (i.e. assuming equal cluster weights as in Sec.~\ref{subsec:equal_weights}), because this is a realistic experimental target on current devices, as evaluated in Sec.~\ref{subsec:exp_results}.

For each Hamiltonian, we found its highest-energy eigenstate by brute force over the $2^m$ basis states, though in a real test, we would approximately optimize with QAOA. This is the solution one would expect to find with Grover's search algorithm, and it can also be interpreted as a bound on the best possible QAOA performance. The highest eigenstate is the weighted MAX-CUT solution or equivalently, the best cluster assignment on the coreset. For the infinite order Hamiltonian, this highest eigenstate is truly the optimal clustering of the coreset, whereas running $2$-means on the coreset does not guarantee an optimal solution. However, note that the optimal clustering on the coreset does not necessarily correspond to the optimal clustering on the whole data set, because the coreset is not completely representative of the underlying data set.

\subsection{Coreset and QAOA Bound Results}\label{subsec:results}

\begin{figure}[t]
\centering
    \centering
    \begin{subfigure}{.95\textwidth}
    \includegraphics[width=\textwidth]{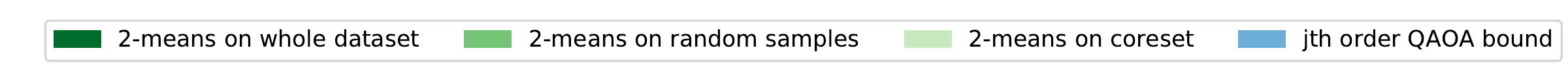}
    \end{subfigure}
    \hfill
    \begin{subfigure}{.49\textwidth}
    \includegraphics[width=\textwidth]{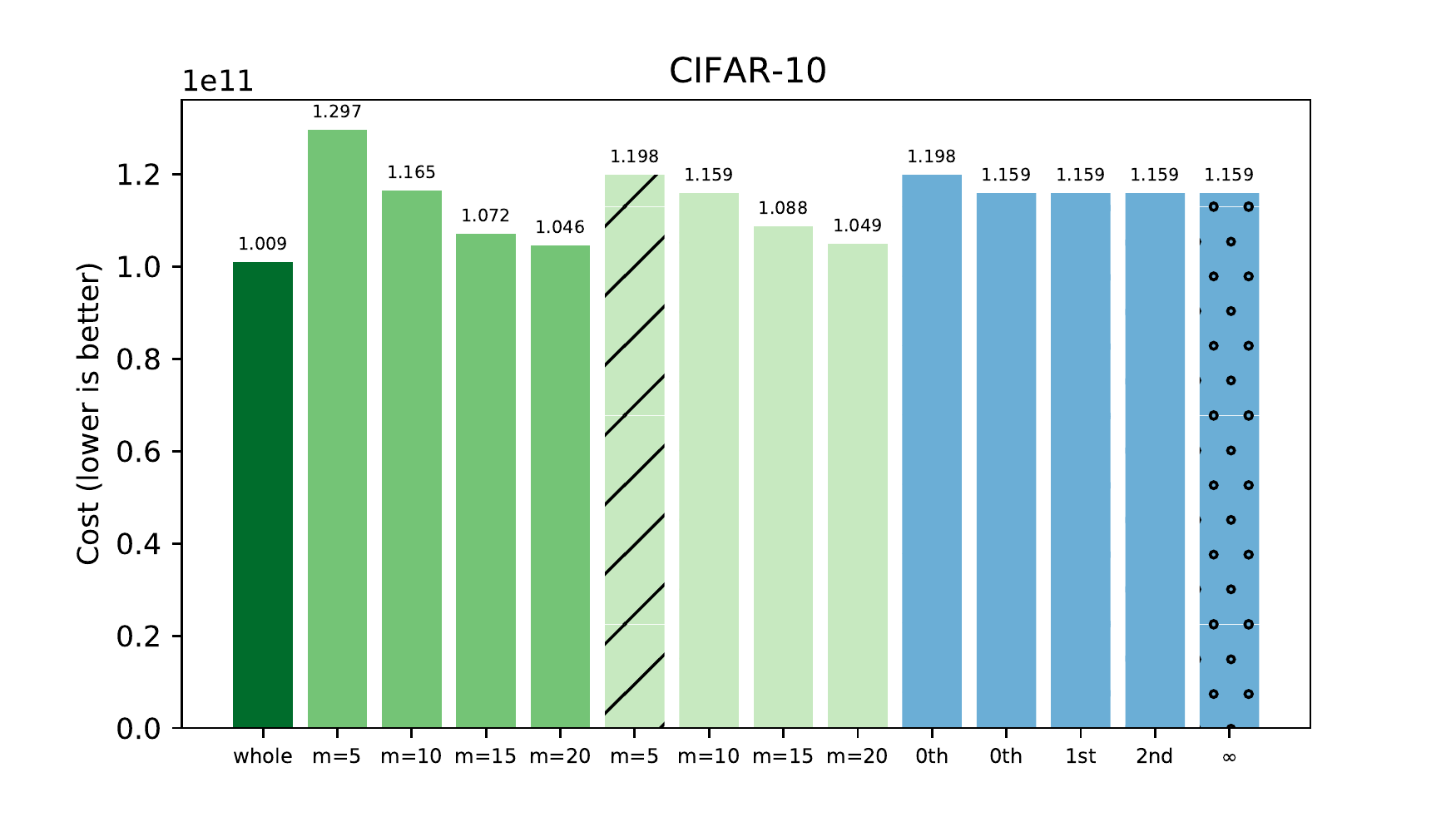}
    \end{subfigure}
    \hfill
    \begin{subfigure}{.49\textwidth}
    \includegraphics[width=\textwidth]{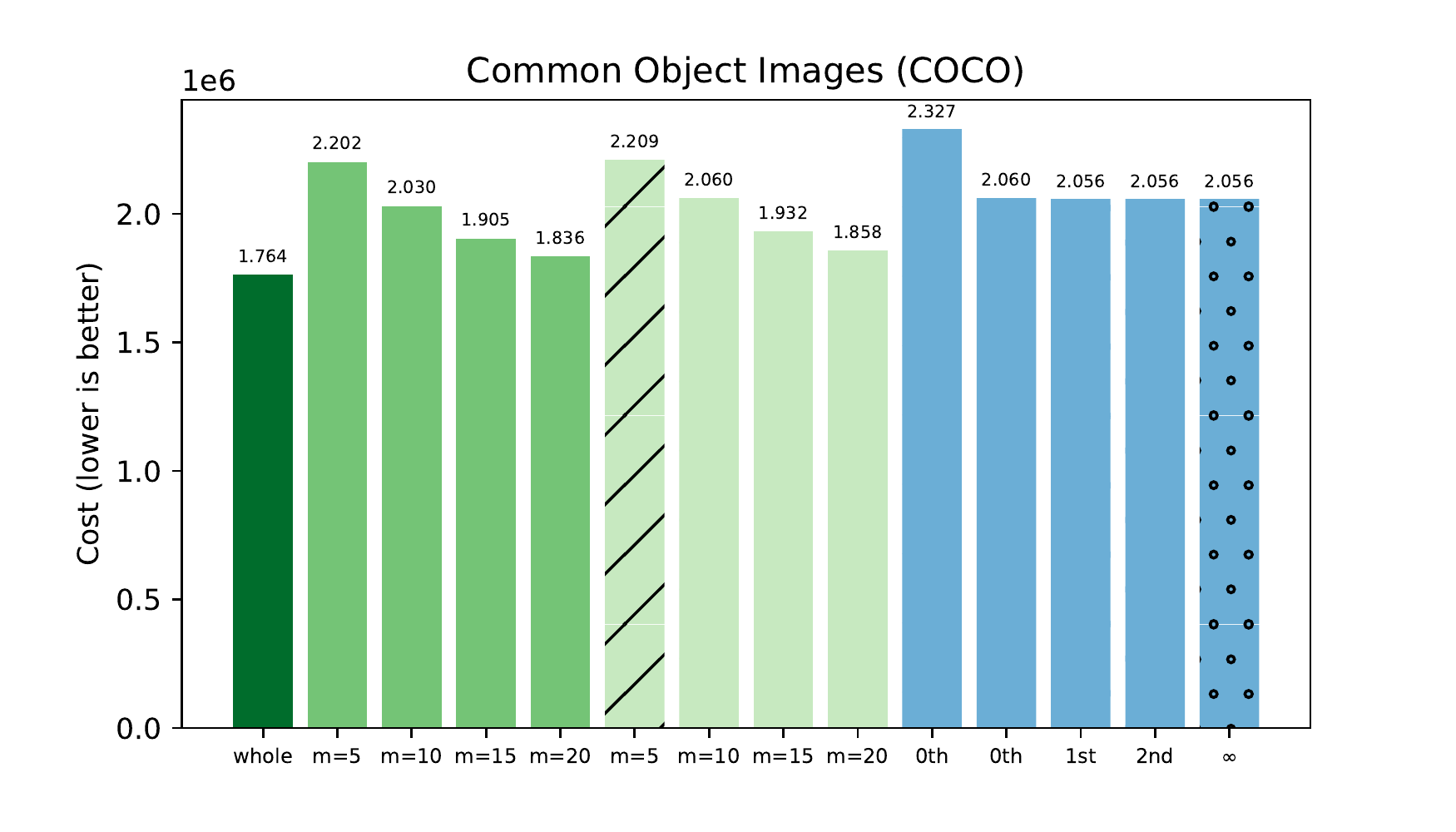}
    \end{subfigure}
    \hfill
    \begin{subfigure}{.49\textwidth}
    \includegraphics[width=\textwidth]{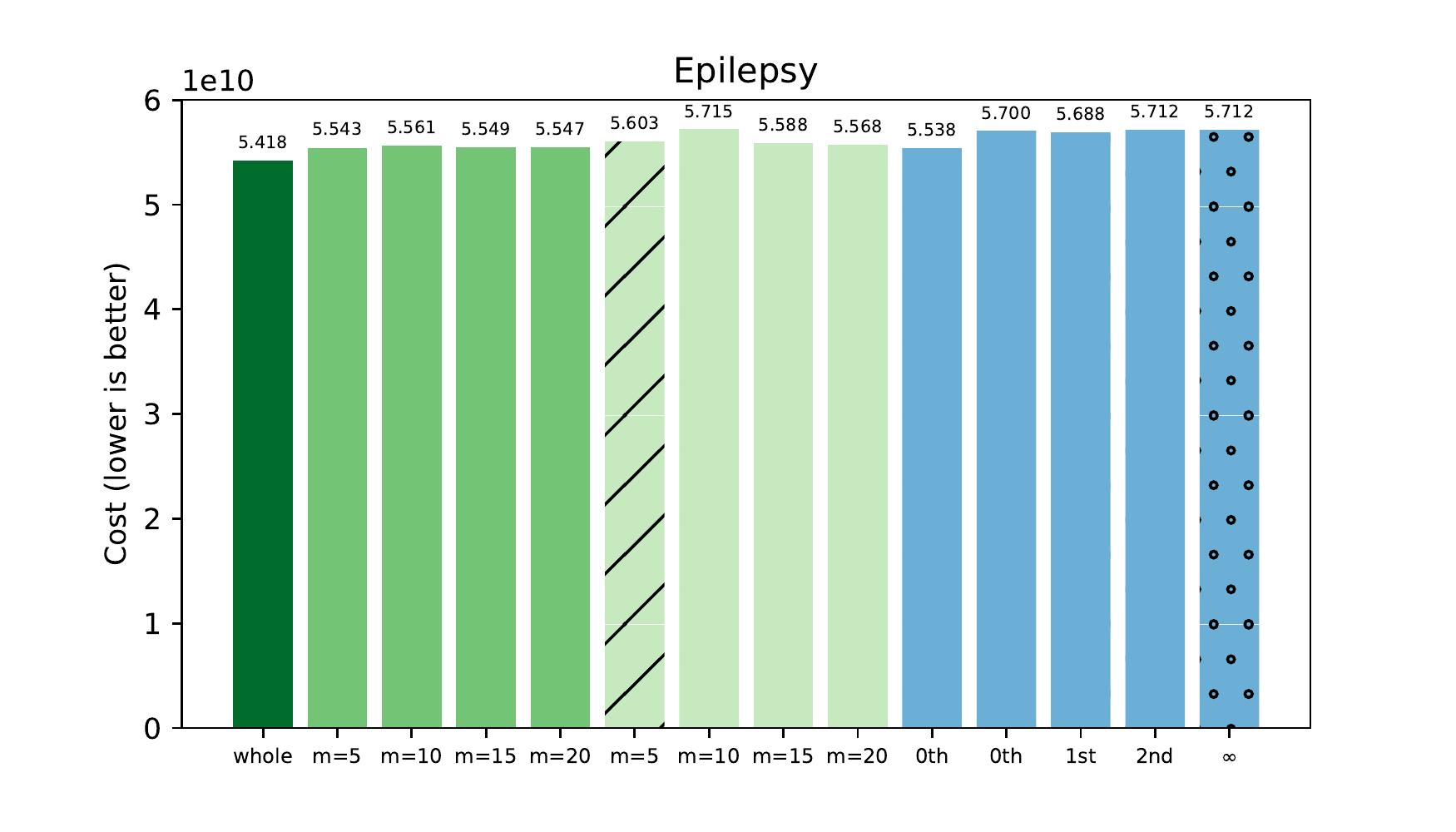}
    \end{subfigure}
    \hfill
    \begin{subfigure}{.49\textwidth}
    \includegraphics[width=\textwidth]{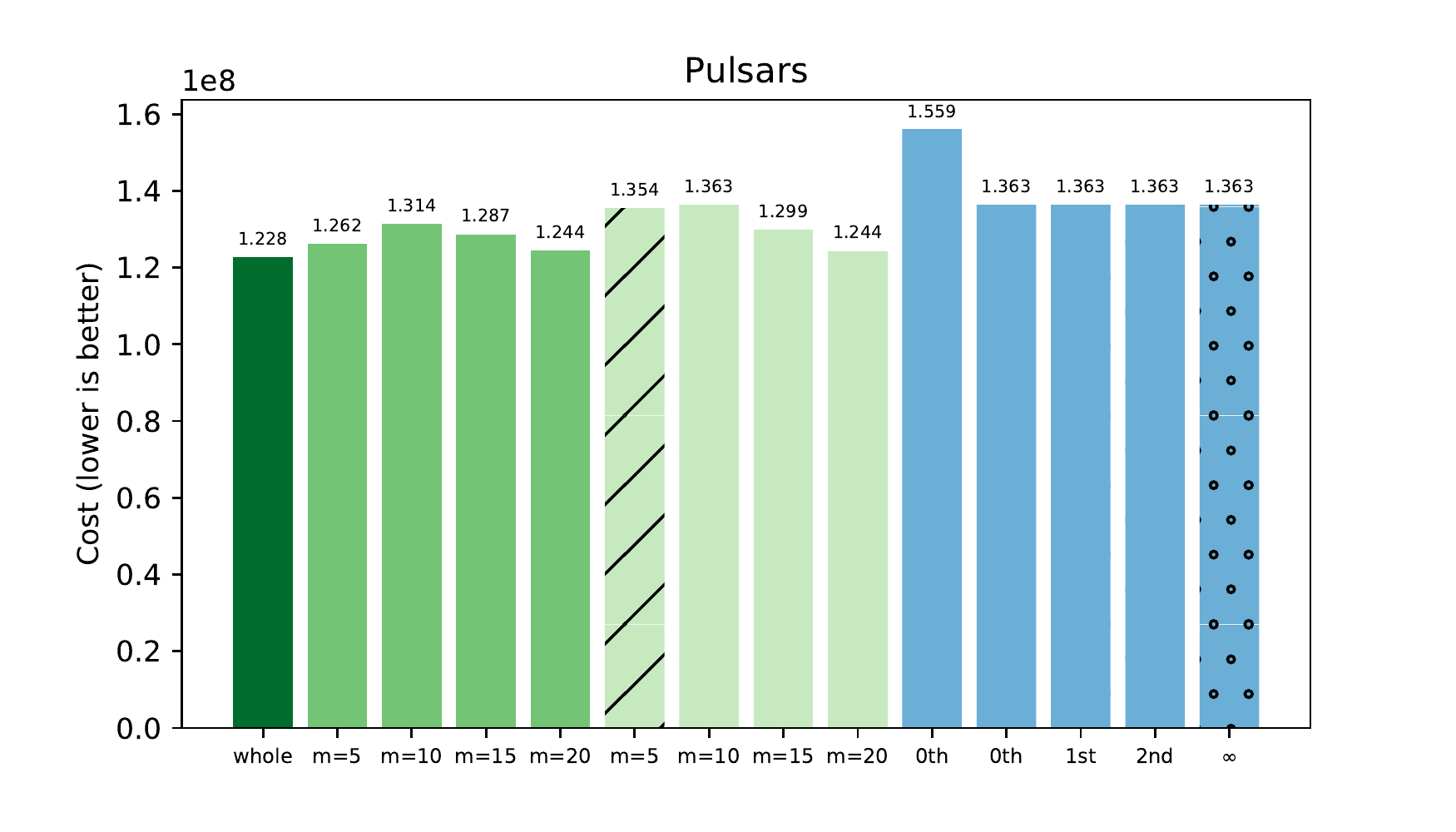}
    \end{subfigure}
    \hfill
    \begin{subfigure}{.49\textwidth}
    \includegraphics[width=\textwidth]{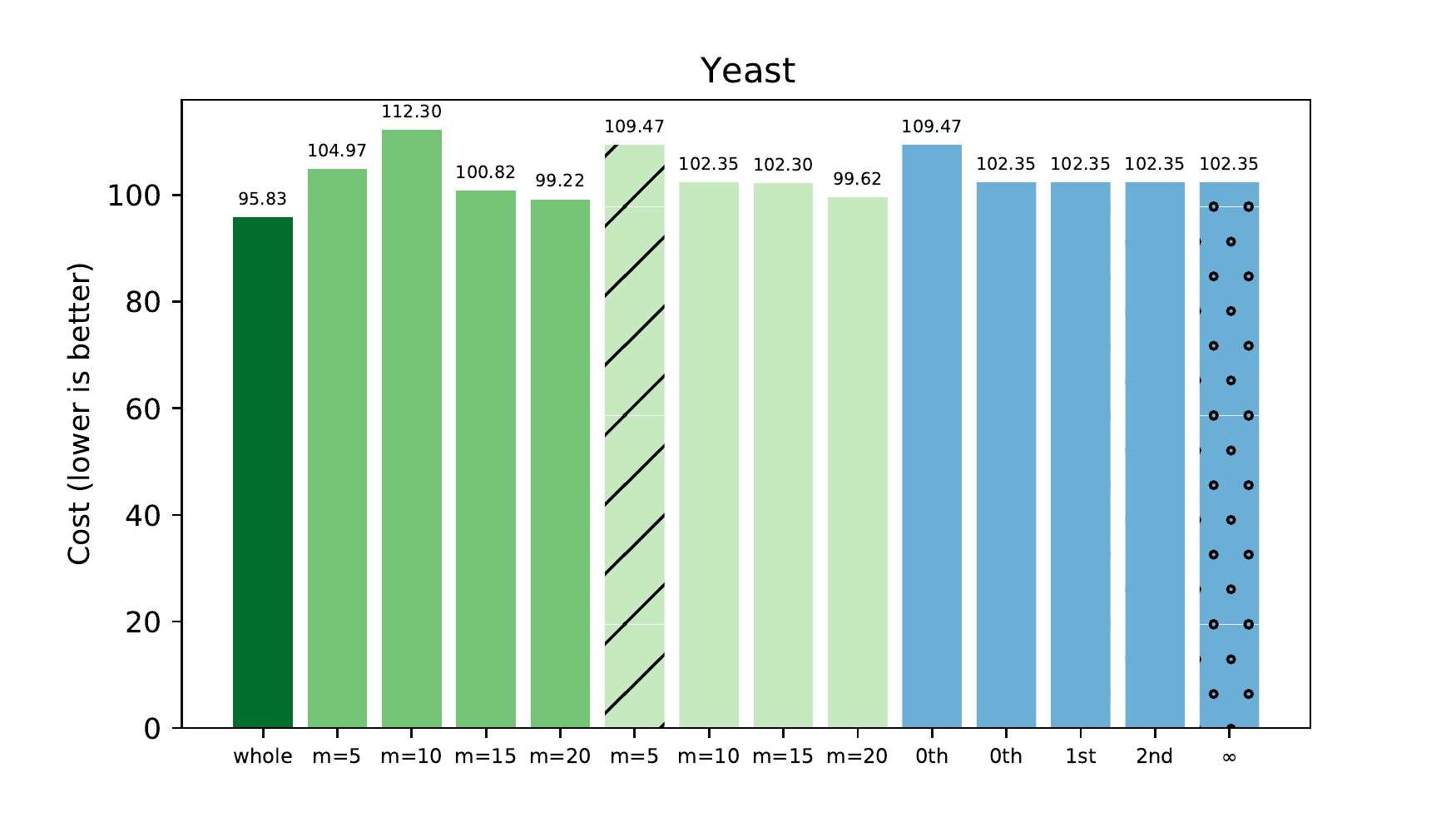}
    \end{subfigure}
    \hfill
    \begin{subfigure}{.49\textwidth}
    \includegraphics[width=\textwidth]{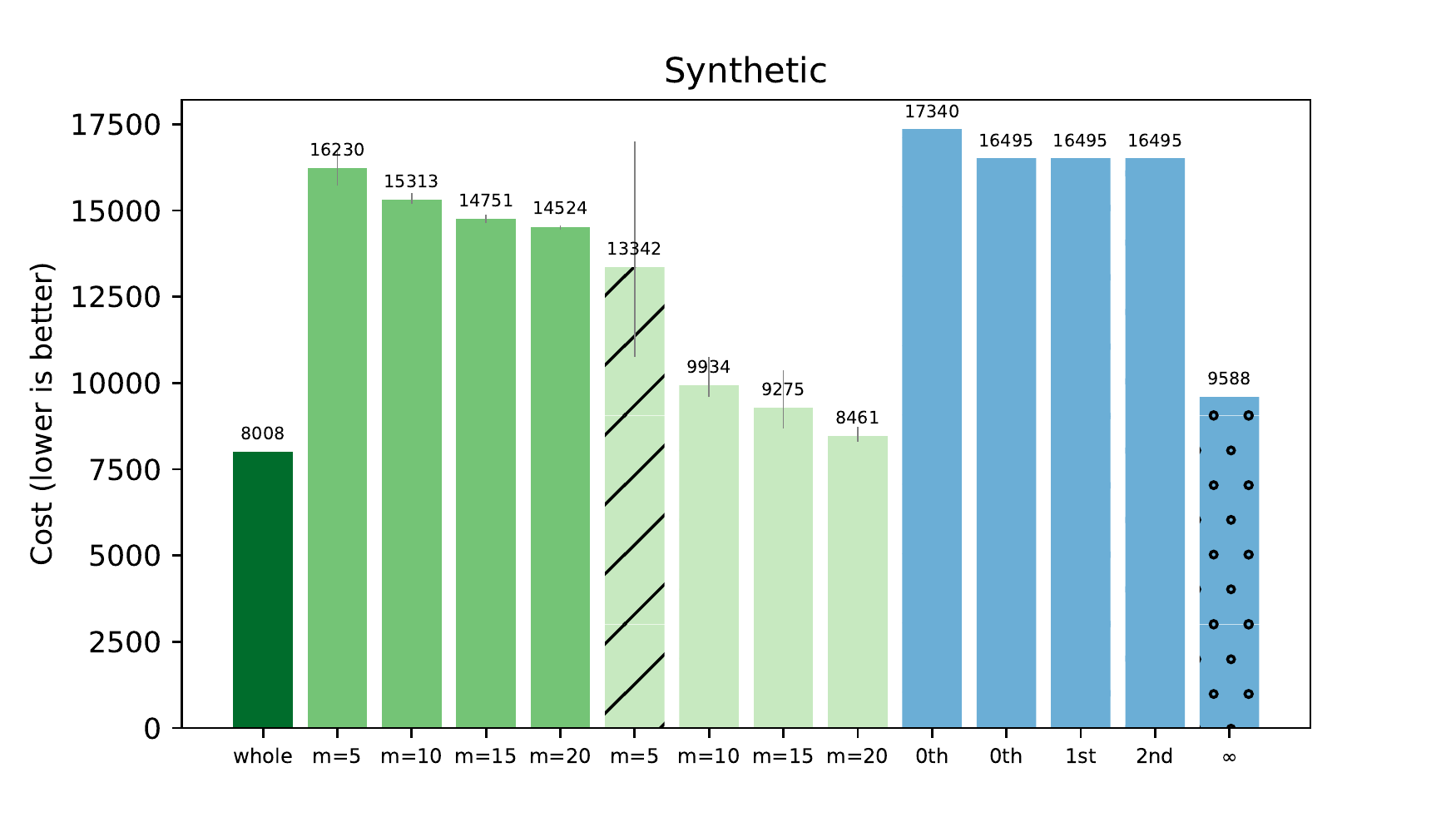}
    \end{subfigure}
\caption{Evaluation on six different data sets. The green bars all result from running 2-means: on the whole data set, and on $m$ = 5, 10, 15, 20 uniformly random samples or coresets. We report the best-of-10 for all data sets, except for Synthetic, which shows the means and min-max error bars. The blue bars express the cost on the whole data set of the highest-eigenstate of the $m=5$ or 10 Hamiltonian using a $j$th order Taylor expansion. The blue bars can be interpreted as a bound on QAOA's performance with $m$ qubits. The texture indicates correspondences between $m=5$ and $m=10$ results.}
\label{fig:megaresults}
\end{figure}

Figure~\ref{fig:megaresults} shows results, using the methodology above, on our six data sets. The green bars, which are entirely classical, correspond to 2-means on the whole data set, a random sample, or a coreset via Algorithm 2 of \cite{braverman2016new}. Interestingly, for most of our benchmarked data sets, \cite{braverman2016new} does not appear to outperform simple random sampling. The exception is on the Synthetic data set, which has 39,950 points in a localized Gaussian cluster, along with 50 points in ten distant clusters. On this data set, random sampling does not perform well, because the random samples are unlikely to capture points outside the big cluster. This suggests we may see gains from using coresets on anomaly-detection oriented data sets.

The blue bars correspond to the energy-maximizing eigenstate of each Hamiltonian, which is constructed from a coreset of $m$ elements and a given Taylor expansion order. QAOA attempts to find this energy-maximizing eigenstate, but it is only an approximate solver. Therefore, the blue bars can be interpreted as a lower bound on the cost of a QAOA-based solution. However, recall that the ultimate cost is with respect to the whole data set, and the coreset is only an approximation of this data set. Therefore, a suboptimal eigenstate of the Hamiltonian could actually outperform the supposedly-optimal eigenstate.

We observe this behavior in a few data sets. Focusing on the blue bars only, we see that the cost is almost always lower (or the same) when we increase $m$ or increase the order of the Hamiltonian. However, for the Epilepsy data set, we see that the dotted ($m=10$) blue bars have lower cost for lower Taylor expansion orders (i.e. more approximate Hamiltonians). In fact, the $m=5$ diagonal-striped blue bar has the lowest cost among blue bars. A possible explanation for this is that the $m=10$ coreset (green dotted bar) is poor, so deviating from its optimal clustering is actually favorable for the overall cost. This does open the broader possibility that QAOA's approximate optimization could outperform brute force optimization, but this may simply be an artifact of the limitations of coresets.

On all six data sets studied, the lowest clustering cost is achieved by running 2-means on the whole data set. The main barrier to a possible `quantum advantage' seems to be that for small $m$, the coreset approximation is too coarse. However, we were able to find QAOA bounds that beat corresponding coreset 2-means results. This occurred in both the Epilepsy and Synthetic data sets, where a blue quantum bar beats the same-textured green bar (although again, this may be an artifact of a poor coreset). It is also encouraging that the 0th order QAOA bounds, which have only quadratic gate count cost, are generally similar to the higher order bounds. The exception is the Synthetic data set, which by construction favors unequal cluster sizes and violates the equal cluster weight assumption of 0th order expansion. The broader pattern appears to be that if a data set is amenable to coresets, it does not work well for low-order QAOA approximation, and vice versa.

\begin{figure}[t!]
    \centering
    \includegraphics[width=1.0\textwidth]{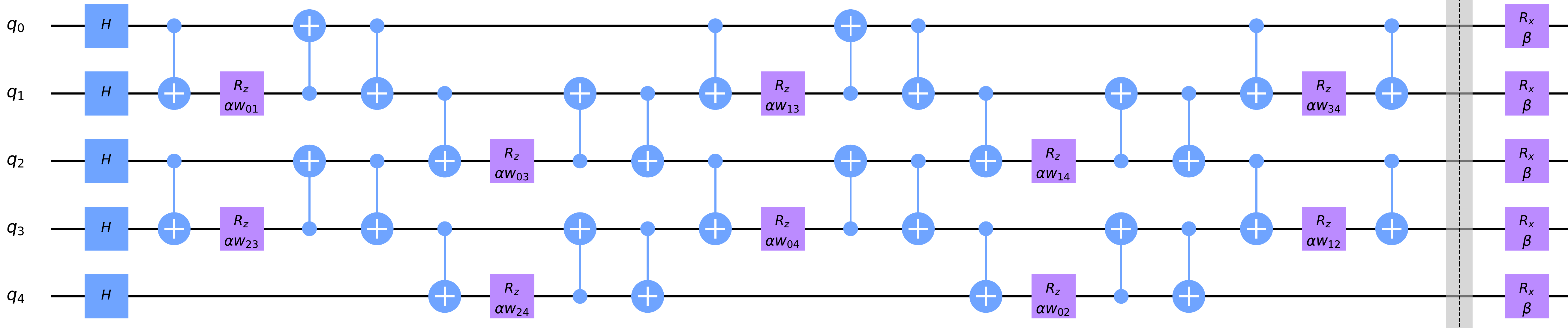}
    \caption{Example of a QAOA circuit for $k$-means on an $m=5$ coreset implemented using the swap networks proposed in~\cite{kivlichan2018quantum, o2019generalized}. Here, $P=1$, $\alpha$ and $\beta$ are the variational parameters and the $w_{ij}$'s are the edge weights of the constructed graph (see Figure~\ref{fig:weighted_maxcut}).}
    \label{fig:qaoa_circuit}
\end{figure}

\subsection{Experimental QAOA Results}\label{subsec:exp_results}
For each of the six data sets examined above we also tested the performance of QAOA for the $m=5$ case on the 5-qubit \textit{ibmq\_rome} processor. Figure~\ref{fig:qaoa_circuit} shows the QAOA circuit. Solving $k$-means via QAOA on the zeroth or first order Hamiltonian is equivalent to finding the weighted MAX-CUT on a complete graph, which requires an interaction between every pair of qubits during each step of QAOA. This would seem to imply that the quantum computer must support a high degree of connectivity between qubits or else incur a large number of SWAP operations which would deteriorate performance. However, using the SWAP networks proposed in~\cite{kivlichan2018quantum, o2019generalized}, we can achieve the required all-to-all interactions in depth which scales linearly with the number of qubits $m$, while only requiring nearest-neighbor interactions. SWAP networks were also implemented for MAX-CUT QAOA in~\cite{crooks2018performance} where the total number of CNOTs required was found to scale as $\frac{3}{2}m(m-1)P$. We note that this CNOT count can be slightly reduced by forgoing the last layer of the SWAP network and reordering the bits in a classical post-processing step so that the overall number of CNOTs scales as $(\frac{3}{2}m(m-1)-\lfloor\frac{m}{2}\rfloor)P$.

\begin{figure}[t]
    \centering
    \includegraphics[width=1.0\textwidth]{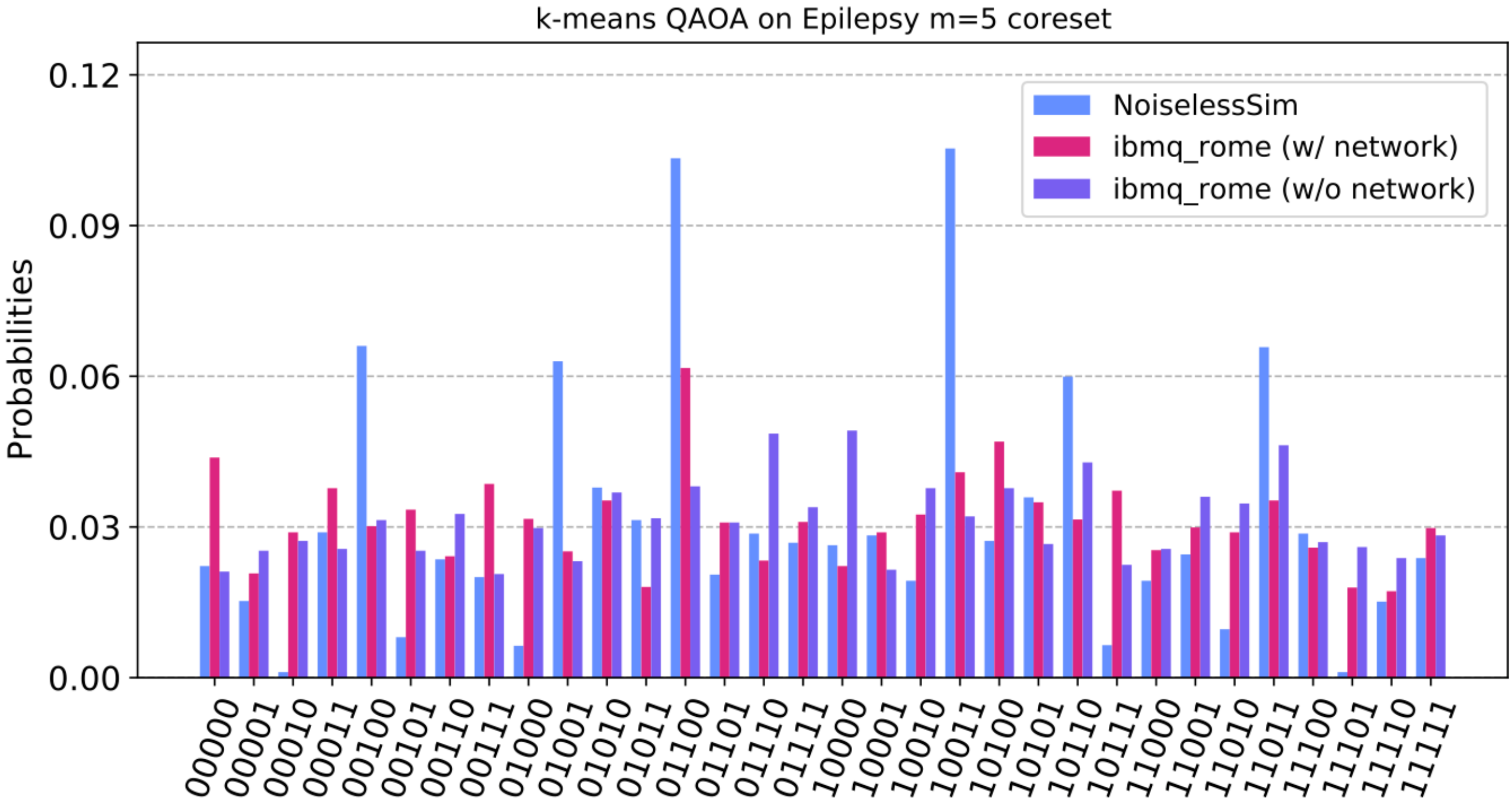}
    \caption{Example of a QAOA circuit for $k$-means on an $m=5$ coreset implemented using the SWAP networks proposed in~\cite{kivlichan2018quantum}. Here, $P=1$, $\alpha$ and $\beta$ are the variational parameters and $w_{ij}$ are the edge weights of the constructed graph (see Figure~\ref{fig:weighted_maxcut}). Each execution consists of 8192 individual shots.}
    \label{fig:epilepsy_results}
\end{figure}

Figure~\ref{fig:epilepsy_results} shows the results of running $k$-means QAOA for the Epilepsy $m=5$ coreset on \textit{ibmq\_rome} with and without the SWAP network. Before running on the quantum hardware, we found optimal values for the variational parameters using circuit simulation in conjunction with a Nelder--Mead optimizer. In Figure~\ref{fig:epilepsy_results} the best partitionings found by the noiseless simulation are the symmetric 01100 and 10011 states. We use the cluster centers given by the coreset partitioning to evaluate the cost function on the entire data set; both the 01100 and 10011 partitions have a cost $c=5.581\mathrm{e}{10}$ which approaches the bound shown in Figure~\ref{fig:megaresults}. There is a significant amount of noise present in the hardware executions compared to the noiseless simulation, but when the SWAP network is utilized, the 01100 solution state is still the most likely state to be measured. Without the SWAP network, the highest probability state would result in a suboptimal partition.

\section{Discussion}\label{sec:discussion}

In this work we investigated the performance of performing $k$-means clustering using QAOA, using offline coresets to effectively reduce the size of the clustering data sets. Indeed, we found that there do exist data sets where coresets seem to work well in practice, such as the Synthetic data set we analyzed in Sec.~\ref{subsec:results}. Furthermore, as our Hamiltonian construction of the problem is all-to-all connected, our QAOA instance circumvents the light cone oriented issues~\cite{farhi2020quantum, bravyi2019obstacles} associated with running constant-$P$-depth QAOA on sparse bounded-degree graphs. Furthermore, while naively implementing QAOA on this all-to-all connected graph would require time evolution under $O(m^2)$ edge operators for a size $m$ instance, the SWAP network construction allows us to implement QAOA evolution with just $O(m)$ depth, even with linear connectivity.

However, in practice coresets did not seem to work well relative to random sampling for the standard classification data sets we benchmarked our results on. This may be due to the small $m$ we restricted our coresets to, with the motivation of fitting the problem instance on today's quantum computers, or may be due to the fact that these ``natural'' data sets have near equally sized optimal clusters. Indeed, the Synthetic data set where using coresets performed well had artificially rare clusters that naive random sampling would miss---however, this worked to the detriment of our Hamiltonian construction of the problem, which involves Taylor expanding the optimization problem around near equal optimal cluster sizes. As standard $k$-means already performs remarkably well, it seems that one would need a high-degree Hamiltonian expansion for a method such as QAOA to compete, or otherwise a more clever Hamiltonian construction of the problem. Methods such as Grover's algorithm, however, would not necessarily have this limitation. We leave for future work refining this intuition and perhaps finding instances where both coresets and QAOA work well in conjunction.

\begin{acknowledgments}
This work is funded in part by EPiQC, an NSF Expedition in Computing, under grants CCF-1730449/1729369/1730082; in part by STAQ under grant NSF Phy-1818914; and in part by DOE grants DE-SC0020289 and DE-SC0020331. P.G. is supported by the Department of Defense (DoD) through the National Defense Science \& Engineering Graduate Fellowship (NDSEG) Program. E.R.A. is supported by the National Science Foundation Graduate Research Fellowship Program under Grant No. 4000063445, and a Lester Wolfe Fellowship and the Henry W. Kendall Fellowship Fund from M.I.T.
\end{acknowledgments}

\bibliographystyle{unsrt}
\bibliography{refs}

\appendix

\section{$2$-Means to Weighted MAX-CUT Reduction} \label{app:reduction}
In Sec.~\ref{subsec:equal_weights}, it is claimed that minimizing Eq.~\eqref{eq:2means_weighted} (the $2$-means objective function on a weighted coreset) over set assignments $S_{\pm 1}$, is equivalent to maximizing Eq.~\eqref{eq:weighted_cluster_distance} (the weighted intercluster distance):
\begin{equation}
\begin{split}
&\argmin_{S_{\pm 1}} \sum_{k \in \{-1, +1\}} \sum_{i \in S_{k}} w_i\left\lVert \bm{x}_i - \bm{\mu}_{k} \right\rVert^2 \\
&= 
    \argmax_{S_{\pm 1}}\left( W_{+1}W_{-1} \left\lVert \bm{\mu}_{+1} - \bm{\mu}_{-1}\right\rVert^2 \right).
\end{split}
\end{equation}

This essence of this reduction is a known~\cite{kriegel2017black} application of the Law of Total Variance. Our notation and proof are drawn directly from~\cite{bauckhage2017adiabatic}, with some deviations to handle weighted $k$-means clustering. We begin by considering the coreset's scatter, which measures the sum of squared distances from weighted coreset points to the overall centroid. The scatter is a function of the coreset data, and is therefore invariant with respect to the chosen cluster partitioning. We will write the scatter as a sum over three terms:
\begin{equation}
\begin{split}
    &\sum_{i} w_i\left\lVert \bm{x}_i - \bm{\mu}\right\rVert^2 \\
    &=\sum_{k \in \{-1, +1\}} \sum_{i \in S_{k}} w_i \left\lVert \bm{x_i} - \bm{\mu} \right\rVert^2 \\
    &= \sum_{k \in \{-1, +1\}} \sum_{i \in S_{k}} w_i \left\lVert\left(\bm{x}_i - \bm{\mu}_{k}\right) - \left(\bm{\mu} - \bm{\mu}_{k}\right)\right\rVert^2 \\
    &= \left(\sum_{k \in \{-1, +1\}} \sum_{i \in S_{k}} w_i \left\lVert\bm{x}_i - \bm{\mu}_{k}\right\rVert^2\right) \\
    &-2 \left(\sum_{k \in \{-1, +1\}} \sum_{i \in S_{k}} w_i \left(\bm{x}_i - \bm{\mu}_{k}\right) \cdot \left(\bm{\mu} - \bm{\mu}_{k}\right)\right) \\
    &+ \left(\sum_{k \in \{-1, +1\}} \sum_{i \in S_{k}} w_i \left\lVert\bm{\mu} - \bm{\mu}_{k} \right\rVert^2\right) \\
    &= (T_1) - 2(T_2) + (T_3) \\
\end{split}
\end{equation}

$T_1$ is the just the 2-means objective from Eq~\eqref{eq:2means_weighted}, which we wish to minimize. Meanwhile, $T_2$ is zero:
\begin{equation}
\begin{split}
    T_2 &= \sum_{k \in \{-1, +1\}} \sum_{i \in S_{k}} w_i \left(\bm{x}_i - \bm{\mu}_{k}\right) \cdot \left(\bm{\mu} - \bm{\mu}_{k}\right) \\
    &= \sum_{k \in \{-1, +1\}} \left[\left( \sum_{i \in S_{k}} w_i \bm{x}_i \right) - W_{k} \bm{\mu}_{k} \right] \cdot \left(\bm{\mu} - \bm{\mu}_{k}\right) \\
    &= \sum_{k \in \{-1, +1\}} \left[ W_{k} \bm{\mu}_{k} - W_{k} \bm{\mu}_{k} \right] \cdot \left(\bm{\mu} - \bm{\mu}_{k}\right) \\
    & = 0.
\end{split}
\end{equation}

Finally, we show that $T_3$ is related to the weighted intercluster distance objective:
\begin{widetext}
\begin{equation*}
\begin{split}
&T_3 = \sum_{i \in S_{-1}} w_i \left\lVert\bm{\mu} - \bm{\mu}_{-1}\right\rVert^2 + \sum_{i \in S_{+1}} w_i \left\lVert \bm{\mu} - \bm{\mu}_{+1}\right\rVert^2 \\
&= W_{-1} \left\lVert\bm{\mu}-\bm{\mu}_{-1}\right\rVert^2 + W_{+1} \left\lVert\bm{\mu} - \bm{\mu}_{+1}\right\rVert^2 \\
&= W_{-1} \left\lVert\frac{W_{-1} \bm{\mu}_{-1} + W_{+1} \bm{\mu}_{+1}}{W}-\bm{\mu}_{-1}\right\rVert^2 + W_{+1} \left\lVert\frac{W_{-1} \bm{\mu}_{-1} + W_{+1} \bm{\mu}_{+1}}{W} - \bm{\mu}_{+1}\right\rVert^2 \\ 
&= W_{-1} \left\lVert \frac{ W_{+1} \bm{\mu}_{+1} - W_{+1} \bm{\mu}_{-1}}{W} \right\rVert^2 + W_{+1} \left\lVert \frac{ W_{-1} \bm{\mu}_{-1} - W_{-1} \bm{\mu}_{+1}}{W} \right\rVert^2 \\
&= \frac{W_{-1} W_{+1}^2}{W^2} \left\lVert\bm{\mu}_{+1} - \bm{\mu}_{-1}\right\rVert^2 + \frac{W_{+1} W_{-1}^2}{W^2} \left\lVert\bm{\mu}_{-1} - \bm{\mu}_{+1}\right\rVert^2 \\
&= \frac{W_{-1} W_{+1}}{W} \left\lVert\bm{\mu}_{-1} - \bm{\mu}_{+1}\right\rVert^2.
\end{split}
\end{equation*}
\end{widetext}

The last line is just the weighted intercluster distance objective (in Eq.~\eqref{eq:weighted_cluster_distance}), scaled by a $\frac{1}{W}$ constant. Since the scatter ($T_1 -2T_2 + T_3 = T_1 + T_3$) is not a function of the partitioning, minimizing $T_1$ corresponds to maximizing $T_3$. Therefore, minimizing the $2$-means objective is equivalent to maximizing the weighted intercluster distance, i.e. maximizing $W_{-1} W_{+1} \left\lVert\bm{\mu_{-1}} - \bm{\mu_{+1}}\right\rVert^2$. As shown in Sec.~\ref{subsec:equal_weights}, this optimization problem corresponds to a weighted MAX-CUT instance.

\end{document}